\documentclass{article}
\usepackage[T1]{fontenc} 
\usepackage[utf8]{inputenc} 
\usepackage{ismir,amsmath,cite,url}
\usepackage{graphicx}
\usepackage{color}

\usepackage{subfig}
\usepackage{multirow}
\usepackage{enumerate} 
\usepackage{makecell}

\usepackage{amsfonts}
\usepackage{xcolor}

\newcommand{\yhyang}[1]{{\color{black}#1}}

\title{DDSP-based Singing Vocoders: A New 
Subtractive-based Synthesizer and A Comprehensive Evaluation}

\multauthor
{Da-Yi Wu$^{1\ast}$  \hspace{1cm} Wen-Yi Hsiao$^{2\ast}$ \hspace{1cm} Fu-Rong Yang$^{3\ast}$ \hspace{1cm} Oscar Friedman$^4$}
{\bfseries{Warren Jackson$^5$ \hspace{1cm} Scott Bruzenak$^4$ \hspace{1cm}Yi-Wen Liu$^{3}$ \hspace{1cm} Yi-Hsuan Yang$^{1,2}$}\\
\\
$^1$~Academia Sinica,  
$^2$~Taiwan AI Labs,   
$^3$~National Tsing Hua Univ.,
$^4$~470 Music Group, 
$^5$~PARC\\
{\tt\small \{ericwudayi2, s101062219, fjbcrs34\}@gmail.com}
}

\def\authorname{Wu \emph{et al.}}
\usepackage[bookmarks=false,hidelinks]{hyperref}
\begin{document}
\maketitle
\renewcommand{\thefootnote}{\fnsymbol{footnote}}
\begin{abstract}
A vocoder is a conditional audio generation model that converts acoustic features such as  mel-spectrograms into waveforms. 
Taking inspiration from
Differentiable Digital Signal Processing (DDSP), we propose a new vocoder named SawSing for singing voices.
SawSing synthesizes the harmonic part of singing voices by filtering a sawtooth source signal with a linear time-variant finite impulse response filter whose coefficients are estimated from the input mel-spectrogram by a neural network.
As this approach enforces phase continuity, SawSing can generate  singing voices without the 
phase-discontinuity glitch of 
many existing 
vocoders. 
Moreover, the source-filter assumption provides an inductive bias that allows SawSing to be trained on a small amount of data.
Our evaluation shows that SawSing converges much faster and outperforms state-of-the-art generative adversarial network- and diffusion-based vocoders in a resource-limited scenario with only 3 training recordings and a 3-hour training time.\footnote{Equal contribution. Preliminary work was done while Wu was a remote intern working with Friedman at 470 Music Group, LLC.}
\end{abstract}

\renewcommand*{\thefootnote}{\arabic{footnote}}
\setcounter{footnote}{0}

\section{Introduction}
\label{sec:intro}

\begin{figure}[t]
\begin{center}
\includegraphics[width=\linewidth]{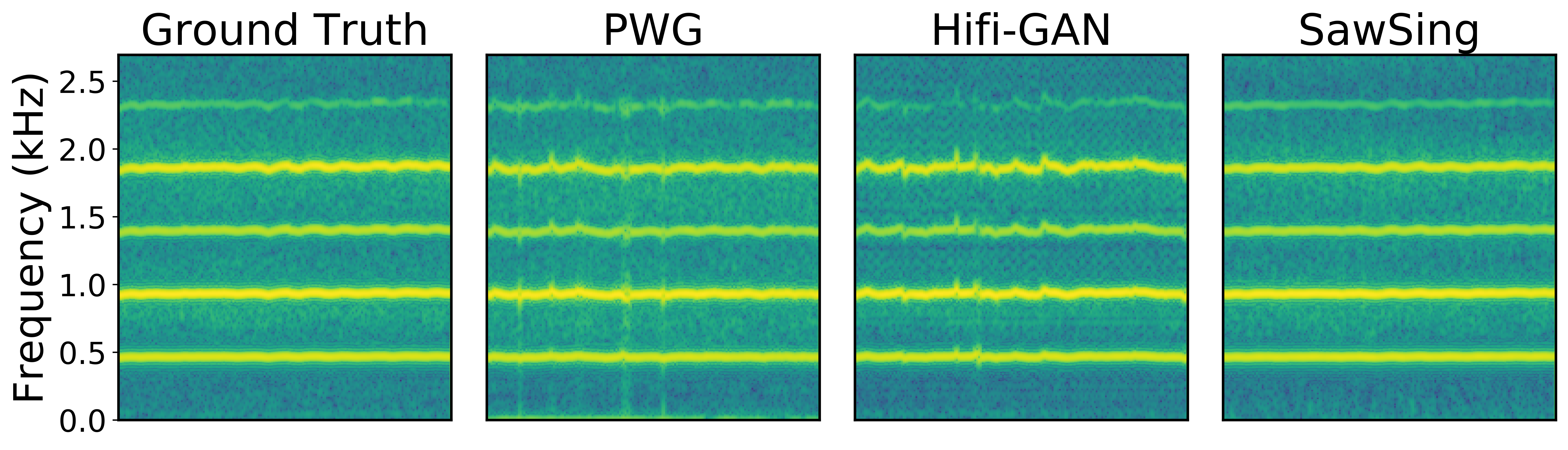}
\end{center}
\vspace*{-3mm}
\caption{The magnitude spectrograms of a long utterance of an original recording (`ground truth') and those reconstructed by two widely-used neural vocoders, Parallel WaveGAN (PWG) \cite{pwg} and HiFi-GAN \cite{hifigan}, and the proposed SawSing. 
Each vocoder is trained on 3 hours of recordings from a female singer until convergence. We see glitches in the results of PWG and HiFi-GAN.}
\label{fig:mel}
\end{figure}

Singing voice synthesis (SVS) aims to generate human-like singing voices from musical scores with lyrics \cite{cook1996singing,lee2019adversarially,9053944,ren2020deepsinger,hifi,sinsy21taslp,survey,karasinger,diff}. 
State-of-the-art (SOTA) voice synthesis techniques involve two stages:  acoustic feature modeling from musical scores and audio sample reconstruction via a so-called ``vocoder.'' A \emph{neural vocoder} takes an acoustic feature such as mel-spectrogram as input and outputs a waveform using deep learning networks \cite{wavenet,wrnn,nsf,melgan,pwg,hifigan,periodnet21icassp,guo22icassp,singgan,chen2021wavegrad,diffwave,fastdiff,lam2022bddm}. However, phase discontinuities within partials often occur due to the difficulty of reconstructing realistic phase information from a mel-spectrogram. This may lead to a 
short-duration broadband transient perceived as ``glitch'' or ``voice tremor,'' which is more audible during long utterances commonly found in singing \cite{singgan}, as exemplified in Figure \ref{fig:mel}.

Differentiable Digital Signal Processing (DDSP) \cite{ddsp} introduces a new paradigm for neural audio synthesis. It incorporates classical digital signal processing (DSP) synthesizers and effects as differentiable functions within a neural network (NN), and combines the expressiveness of an NN with the interpretability of classical DSP. The use of phase-continuous oscillators is a potential solution to the phase problem from which regular neural vocoders suffer, and the strong inductive bias of this approach may obviate the need of large training data. 
Furthermore, DDSP has already succeeded in achieving
sound synthesis of, and timbre transfer between, monophonic instruments \cite{ddsp,granule,NNwaveshaping,diffsynth,dwts,ddsp-iui,ganis21ddsp,nercessian21waspaa}.
These motivate us to explore whether the DDSP approach can be applied to build a singing vocoder.



This paper proposes SawSing, a DDSP-based singing vocoder which reconstructs a monophonic singing voice from a mel-spectrogram. 
The architecture of SawSing similarly consists of an NN and classical DSP components; unlike DDSP, its DSP portion is a subtractive harmonic synthesizer which filters a sawtooth waveform containing all possible harmonic partials, plus a subtractive noise synthesizer which filters uniform noise. The sawtooth signal enforces phase continuity \emph{within} partials, thereby avoiding the glitches. 
Moreover, the partials of a sawtooth signal are guaranteed to be in phase, so it also enforces the phase coherence \emph{between} partials, intrinsic to human voices. 
The function of the NN, on the other hand, is to infer from the  mel-spectrogram the fundamental frequency (f0) of the sawtooth signal and the filter coefficients of the harmonic and noise synthesizers for each time frame.





In our experiments, we use  
data from two singers (each three hours) of the MPop600 Mandarin singing corpus \cite{mpop600}.
We compare the performance of SawSing with the neural source-filter (NSF)  model \cite{nsf}, two existing DDSP-based synthesizers, the original additive-based DDSP \cite{ddsp} and the differentiable wavetable synthesizer \cite{dwts}, and a few famous neural vocoders, i.e., two generative adversarial network (GAN)-based models  \cite{pwg,hifi}  and a  diffusion-based model \cite{fastdiff}. 
We consider both a regular scenario where the vocoders are trained 
for days
using the 3-hour dataset, and a resource-limited scenario with constraints on training data and training time.
Our experiments show that SawSing 
converges much faster and outperforms the other vocoders in the resource-limited scenario.

The main contribution of the paper is two-fold. First, we show that despite differences between instrumental sounds and singing voices \cite{sundberg89book}, 
the classic idea of subtractive synthesis \cite{lane97,Huovilainen2005NewAT} can be applied to singing voices using the DDSP approach.\footnote{We note that the use of a sawtooth waveform in DDSP-based models has  been attempted for speech synthesis \cite{NHV} 
and instrumental synthesizer sound matching \cite{diffsynth}, but its application to singing vocoder is new.} 
Second, we provide empirical evidences showing that DDSP-based vocoders can compare favorably with sophisticated, SOTA neural vocoders.
Furthermore, since DDSP-based vocoders are lightweight and training-efficient, they have the potential to be used in creative and real-time scenarios of singing expression with limited training data of a target singing voice \cite{greshler2021catch,nercessian20vc}.

We open source our code at { \url{https://github.com/YatingMusic/ddsp-singing-vocoders/}}.
For audio examples, visit our demo webpage { \url{https://ddspvocoder.github.io/ismir-demo/}}. 

\section{Background}
\label{sec:bg}

Neural vocoders often aim to reconstruct a waveform $\mathbf{y}\in \mathbb{R}^{1 \times T}$
from an input mel-spectrogram $\mathbf{X}\in \mathbb{R}^{M \times N}$:
\begin{equation}
\label{eq:vocoder-nn}
   \mathbf{y} = f_{ {\rm vocoder} }( \mathbf{X} ) \,,
\end{equation}
where $M, N, T$ denote respectively the number of mel filter banks, spectral frames, and time-domain samples.
The conversion from $\mathbf{X}$ to $\mathbf{y}$ can be done, for example, by upsampling $\mathbf{X}$ multiple times through transposed \emph{convolutions} until the length of the output sequence matches the temporal resolution of the raw waveform \cite{melgan,hifigan}. As usual reconstruction loss functions such as mean-square errors cannot reflect the perceptual quality of the reconstruction, GAN-based approaches \cite{pwg,melgan,hifigan} learn discriminators to better guide the learning process of the generator (i.e., $f_{{\rm vocoder}}$).
Newer diffusion-based approaches \cite{diffwave,fastdiff} avoid the use of discriminators and learn to convert white Gaussian noises $\mathbf{z}\in \mathbb{R}^{1 \times T}$ (i.e., of the same length as $\mathbf{y}$) into structured waveform $\mathbf{y}$ through a denoising-like Markov chain, using $\mathbf{X}$ as a condition.
The mapping process between $\mathbf{X}$ and $\mathbf{y}$ of such neural vocoders appears to be a black box that is hard to interpret. However, given sufficient training data (e.g., recordings amounting to 24 hours \cite{diffwave,fastdiff,hifigan,ljspeech} or 80 hours \cite{singgan}) and training time (e.g., days), 
SOTA neural vocoders can reconstruct the waveforms with high fidelity.

The majority of neural vocoders, however, have been originally developed for speech. 
When the rate of utterances is fast, as is common in speech, the glitches resulting from the phase discontinuities within partials may be perceptually masked by the natural transients of the voice. However, 
during singing, where long utterances are common, these discontinuities are  more audible.

To improve the performance of GAN-based vocoders for singing voices, the idea of incorporating the f0 information has been explored recently. 
PeriodNet \cite{periodnet21icassp} uses 
f0 
to create sine excitation as input to 
Parallel WaveGAN (PWG) \cite{pwg} to model the periodic part of human voices.
Guo \emph{et al.} \cite{guo22icassp} 
further filter 
such an f0-driven excitation signal with a linear time-variant finite impulse response (LTV-FIR) filter whose coefficients are estimated from
the input mel-spectrogram, and use the resulting ``harmonic signal'' as input to PWG and MelGAN \cite{melgan}. 
SingGAN \cite{singgan} uses more complicated ``adaptive feature learning'' layers to incorporate the f0. 
These models were shown to outperform older GAN-based vocoders such as PWG and MelGAN in listening tests, but no evaluations against the newest GAN-based vocoder HiFi-GAN \cite{hifigan} and diffusion-based vocoders were reported. Moreover, their evaluation did not consider resource-limited scenarios.


We propose in this paper a radically different approach that uses traditional DSP synthesizers (instead of upsampling convolutions) as the backbone for $f_{{\rm vocoder}}$.  
While the ideas in DDSP have flourished and been applied to synthesizing not only instrumental sounds \cite{ddsp,granule,NNwaveshaping,diffsynth,dwts}, but also audio effects \cite{ddsp_iir,ddsp_blackbox,ddsp_mixing,ddsp_distortion,djtransgan}, 
their application to singing synthesis remains under-explored.
The only exception, to our knowledge, is the preliminary work presented by Alonso and Erkut \cite{svs_ddsp}, which employed exactly the same additive synthesizer as the original DDSP paper \cite{ddsp}. 
However, they did not compare the performance of their vocoder with any other vocoders.
Our work extends theirs by using a subtractive harmonic synthesizer instead, with comprehensive performance evaluations against SOTA neural vocoders such as HiFi-GAN and FastDiff \cite{fastdiff}.

We note that, while a DDSP-based vocoder may solve the glitch problems by inducing continuous phase hypothesis using a harmonic synthesizer, this hypothesis may constrain the model learning ability. 
Experiments reported in this paper are needed to study its performance.

Publicly-available training corpora for singing tend to be much smaller than those for speech \cite{mpop600,opencpop} (often $\le$10 hours).
Therefore, besides tackling the glitch problem, 
our premise is that SawSing can learn faster than prevalent neural vocoders without a large training corpus, due to its strong inductive bias.
Moreover, the success of SawSing may pave the way for the exploration of other advanced DSP components for singing synthesis in the future.




\section{Original DDSP-Add Synthesizer}
\label{sec:ddsp}



The idea of DDSP is to use DSP synthesizers to synthesize the target audio, with the parameters of the synthesizers $\Phi$ inferred from the  mel-spectrogram with an NN. Namely,
\begin{equation}
\label{eq:vocoder-ddsp}
   \mathbf{y} = f_{ {\rm DSP} }( \Phi ) \,, \quad 
   \Phi = f_{ {\rm NN} }( \mathbf{X} ) \,.
\end{equation}
The original DDSP model 
\cite{ddsp}, referred to as 
\textbf{DDSP-Add} below, 
adopts the \emph{harmonic-plus-noise} model for synthesis \cite{hpn} and decomposes a monophonic sound into a periodic (harmonic) component $\mathbf{y}_\texttt{h}$ and a stochastic (noise) component $\mathbf{y}_\texttt{n}$, i.e., $\mathbf{y} = \mathbf{y}_\texttt{h} + \mathbf{y}_\texttt{n}$, and reconstructs  them separately with an \emph{additive} harmonic oscillator (thus the name ``-Add'') and a \emph{subtractive} noise synthesizer. \footnote{The terms ``additive'' and ``subtractive'' are used to describe how a signal is synthesized. An additive synthesizer generates sounds by combining multiple sources such as oscillators or wavetables, while a subtractive synthesizer creates sounds by using filters to shape a source signal, typically with rich harmonics, such as a square or sawtooth wave \cite{hpn}.} 
The former computes $\mathbf{y}_\texttt{h}$ as a weighted sum of  $K$ sinusoids corresponding to the f0 and its integer multiples up to the 
Nyquist frequency (for anti-aliasing), for $t \in [1,T]$:
\begin{equation}
    \mathbf{y}_\texttt{h}^{\text{DDSP-Add}}(t) =   A(t)\sum_{k=1}^{K} c_k(t)\,  \sin(\phi_k(t))\,, \label{eq:ddspadd-harmonic}
\end{equation} 
where 
$A(t)$ is the global amplitude corresponding to the time step $t$, $c_k(t)$ is the amplitude of the $k$-th harmonic satisfying $\sum_{k=1}^{K} c_k(t) = 1, c_k(t) \geq 0$, and the instantaneous phase $\phi_k (t)$ is computed by integrating the instantaneous frequency $k f_0(t)$, i.e., 
   $ \phi_k(t) = 2\pi \sum_{\tau=0}^{t} k f_0(\tau) + \phi_{0, k} $, 
with $\phi_{0, k}$ initial phase, set to zero. 
The parameters $A,c_k,f_0$ are estimated by $f_\text{NN}$ for each  frame $i\in [1,N]$ and then upsampled to the time-domain with linear interpolation.
On the other hand, $\mathbf{y}_\texttt{n}$ is obtained by convolving a uniform noise signal $\mathbf{\zeta}$ ranging from $-1$ to $1$ (with the same length as a frame) 
with an LTV-FIR filter
$\psi_\texttt{n}(i) \in \mathbb{R}^{L_\texttt{n}}$
estimated per frame:
\begin{equation} 
    \bar{\mathbf{y}}_\texttt{n}(i) =  \mathbf{\zeta} \ast \psi_\texttt{n}(i)\,. \label{eq:ddspadd-noise}
\end{equation} 
The final $\mathbf{y}_\texttt{n}$ is obtained by overlap-adding sequence of segments 
$\bar{\mathbf{y}}_\texttt{n}(i)$ for the frames $i=1\dots N$.
Jointly, the parameters 
$\Phi := \{A(i),\{c_k(i)\}_{k=1}^K,f_0(i),\psi_\texttt{n}(i)\}_{i=1}^N$
are estimated from the mel-spectrogram $\mathbf{X}$ per frame by $f_\text{NN}$, which is a small network with few parameters.


Engel \emph{et al.}\,\cite{ddsp} showed that DDSP-Add can synthesize realistic violin sounds with only 13 minutes of expressive solo violin performances as training data.
Alonso and Erkut \cite{svs_ddsp} employed DDSP-Add for singing synthesis, but with limited performance evaluation.

\section{Proposed SawSing Vocoder}
\label{sec:method}


Under the same harmonic-plus-noise signal model \cite{hpn},
SawSing modifies the the harmonic synthesizer of DDSP-Add \cite{ddsp} with two ideas. 
First, given the f0 estimated from $\mathbf{X}$, SawSing approximates $\mathbf{y}_\texttt{h}$ by a sawtooth signal, which contains an equal number of even and odd harmonics with decaying magnitudes, dropping the coefficients $A$ and $c_k$:
\begin{equation}
\label{eq:saw}
    \widetilde{\mathbf{y}_\texttt{h}}^{\text{SawSing}}(t) =  \sum_{k=1}^{K}\frac{1}{k}  \sin(\phi_k(t)) \,. 
\end{equation}
Second, $\widetilde{\mathbf{y}_\texttt{h}}$ is treated as the ``excitation signal'' and shaped into the desirable $\mathbf{y}_\texttt{h}$ by means of an LTV-FIR filter $\psi_\texttt{h}(i) \in \mathbb{R}^{L_\texttt{h}}$ (that is different from $\psi_\texttt{n}(i)$).
To apply the filter, we extract the segment of $\widetilde{\mathbf{y}_\texttt{h}}$ corresponding to the same frame $i$
and multiply its short-time Fourier Transform (STFT) element-wise with the STFT of $\psi_\texttt{h}(i)$ in the frequency domain, before converting it back to the time domain with the inverse STFT and overlap-adding.
SawSing uses the same subtractive noise synthesizer as DDSP-Add. Therefore, the parameters to be estimated from $\mathbf{X}$ by $f_\text{NN}$ are $\Phi^{\text{SawSing}} := \{f_0(i),\psi_\texttt{h}(i),\psi_\texttt{n}(i)\}_{i=1}^N$.


We observe that to compute $\mathbf{y}_\texttt{h}$, DDSP-Add learns NN to attenuate each of the $k$ source harmonics \emph{individually} (i.e., with $c_k$), while SawSing entails a \emph{source-filter} model \cite{nsf}, using the f0-constrained sawtooth signal in Eqn.~(\ref{eq:saw}) as the excitation  source and a time-varying filter $\psi_\texttt{h}(i)$ decided by the NN  for spectral filtering.
The filter coefficients correspond to formants produced by the vocal folds and do not correlate with  f0.

Besides differences in the harmonic synthesizer, Saw-Sing also uses a different loss function from DDSP-Add.
For monophonic instrumental sounds, Engel \emph{et al.}\,\cite{ddsp} showed it effective to use the multi-resolution STFT (MSSTFT) loss as the reconstruction loss 
for training. This loss considers the difference between the magnitude spectrograms of the target and synthesized audio, denoted as $\mathbf{S}_j$ and $\widehat{\mathbf{S}_j}$ below, for $J$ different resolutions. 
\begin{equation}
\label{eq:loss_msstft}
    l_{\text{MSSTFT}} = \sum_{j=1}^{J} \|\mathbf{S}_j - \widehat{\mathbf{S}_j}\|_1 + \| \log(\mathbf{S}_j) - \log(\widehat{\mathbf{S}_j}) \|_1 \,.
\end{equation}
For singing voices, however, we found 
that  MSSTFT loss alone cannot train adequately.  
We introduce an additional f0-related loss term to facilitate learning:
\begin{equation}
\label{eq:loss_f0}
    l_{f_0} = \| \log(f_0) - \log(\widehat{f_0}) \|_1 \,,
\end{equation}
where the target f0 (${f_0}$) and the estimated one ($\widehat{f_0}$) are both extracted by the WORLD vocoder \cite{worldvocoder}.
Thus, our Sawsing loss function becomes $l_{\text{total}} = l_{\text{MSSTFT}} + l_{f_0}$.
Moreover, we found 
that training is unstable
unless the gradients between $f_\text{DSP}$  and the head of $f_\text{NN}$ for f0 prediction are detached.

\subsection{Implementation Details}

First, we resampled the audio recordings to 24 kHz and quantized them to 16 bits. 
Next we cropped the recordings into 2-second excerpts (i.e., $T=\text{48k}$) and extracted 80-band mel-spectrograms from each ($M=80$), 
with a Hann window of 1024 samples for STFT and a hop size of 240 samples (i.e., 10ms). Accordingly, we set $N=200$. 

We used filter length $L_\texttt{h}=256$ for the harmonic synthesizer for SawSing, and filter length $L_\texttt{n}=80$ for the subtractive noise synthesizers. 
We used at most $K=150$ sinusoids for SawSing. 
To avoid sound clipping, we applied a global scaling factor of 0.4 to the sawtooth signal in Eqn.~(\ref{eq:saw})  to ensure that the range of the summed sinusoids always lies in $[-1,1]$.



We chose a lite version of the Conformer architecture for $f_\text{NN}$ \cite{conformer},
for its well-demonstrated effectiveness in capturing both local and global information in a sequence of acoustic features in speech tasks.
It consists of a pre-net (shallow 1D convolution with ReLU activation and group normalization),
a self-attention stack (3 layers), 
a convolution stack (2 layers)
with post layer normalization, 
and a final linear layer whose output dimension is equal to the number of synthesis coefficients. 
We used the Adam optimizer with 0.002 learning rate.


While the original DDSP-Add paper \cite{ddsp} uses $J=6$ for MSSTFT, we 
found setting $J=4$ to be sufficient in our implementation. Specifically, we used four different FFT sizes (128, 256, 512, 1024) with 75\% overlapping  among adjacent frames. 
While it is possible to introduce a scaling factor to control the balance between $l_{\text{MSSTFT}}$ and $l_{f_0}$, we found doing so does not markedly improve the result.

\begin{table*}[t]
\centering
    \begin{tabular}{|lcc|cccc|cccc|cccc|}
    \hline
        & & &
        \multicolumn{4}{c|}{MSSTFT\,$\downarrow$}  & \multicolumn{4}{c|}{MAE-f0\,(cent)\,$\downarrow$} 
        & \multicolumn{4}{c|}{FAD\,$\downarrow$} \\
    
        Model & Para-  & RTF &\multicolumn{2}{c}{Female} &\multicolumn{2}{c|}{Male} &\multicolumn{2}{c}{Female} &\multicolumn{2}{c|}{Male} &\multicolumn{2}{c}{Female} &\multicolumn{2}{c|}{Male} \\
        
        & meters &  & (a)  & (b) & (a)  & (b)  & (a)  & (b) & (a)  & (b) & (a)  & (b)  & (a)  & (b)  \\
        \hline
    FastDiff \cite{fastdiff} &15.3M 
    & 0.017   &14.5 &17.9 &11.1 &16.9 &\underline{31} &110 &48 &131 &2.29 &7.40 &3.53 &10.0 \\
    HiFi-GAN \cite{hifigan} &13.9M 
    &0.004 &\underline{7.13} &16.7 &\underline{7.82} &18.9 & 34 &247 &34 &433 &0.59 &3.50 &0.51 &10.5 \\
    PWG \cite{pwg}     &~\,1.5M 
    &0.007 &7.39 &13.0 &7.83 &14.8 &35 &129 &\underline{29} &126 &\underline{0.36} &6.15 &2.56 &6.29  \\
    \hline
    NSF \cite{nsf}     & ~\,1.2M  
    &0.006   &7.51 &10.9     &10.2 &13.4     &37 &~\textbf{50}     &30 & ~\underline{82}     &0.49    &3.73     &2.08    &4.83 \\
    DDSP-Add \cite{svs_ddsp}  & ~\,0.5M 
    &0.003 &7.61 &\underline{9.29} &8.37 &\underline{12.1} &\textbf{28} & ~\underline{70} & \textbf{24} &~\textbf{80} &0.56 &\underline{0.92} &1.06 &\underline{2.09}\\
    DWTS \cite{dwts}     & ~\,0.5M 
    &0.019 &7.72 &9.75 &8.83 &13.0 &\textbf{28} &127 &\textbf{24} &662 &0.60 &2.98 &\underline{0.36} &8.58\\
    \hline
    SawSing  & ~\,0.5M 
    &0.003 &\textbf{6.93} &\textbf{8.79} &\textbf{7.76} &\textbf{11.7} &32 &~76 &30 &~\textbf{80} &\textbf{0.12} &\textbf{0.38} &\textbf{0.22} &\textbf{0.59} \\
        \hline
    \end{tabular}
    
    
    \caption{Objective evaluation results of three existing neural vocoders (the first three), three existing DDSP-based vocoders (middle) and the proposed SawSing vocoder, trained on either a female or a male singer, in either (a)  \emph{regular} scenario [3h data, well-trained] or (b) \emph{resource-limited} scenario [3min data, 3h training]. RTF stands for real-time factor (the inference time in seconds for a one-second excerpt). In each column, we highlight the best result in bold, the second best underlined.}
    \label{tab:obj}
\end{table*}

\section{Experimental Setup}
\label{sec:exp}

\subsection{Baselines}


Our evaluation considers in total six baselines. The first three explicitly employ f0, while the last three do not.

First, we adopted two existing DDSP-based vocoders, the original additive-based DDSP (\textbf{DDSP-Add}) \cite{ddsp,svs_ddsp} 
and the differentiable wavetable synthesizer (DWTS) \cite{dwts}. 
\textbf{DWTS} replaces the fixed sinusoids in the additive 
harmonic synthesizer of DDSP-Add by $K'$ learnable 
(rather than pre-defined) 
one-cycle waveforms (``the wavetables'') 
$\mathbf{w}_k \in \mathbb{R}^{B}, k \in [1,K']$, to gain flexibility to model a wider variety of sounds (but only tested on instrumental sounds in \cite{dwts}). 
Mathematically, 
    $\mathbf{y}^{\text{DWTS}}_\texttt{h}(t) = A(t)\sum_{k=1}^{K'} c_k(t) \sigma(\mathbf{w}_k, \phi_\pi(t))$,
where $\sigma$ is an indexing function that returns a sample of $\mathbf{w}_k$ according to the instantaneous modulo phase $\phi_\pi(t)$ computed from $f_0(t)$.
For fair comparison, we used the same Conformer-like architecture for the $f_\text{NN}$ for DDSP-Add, DWTS, and SawSing, and the same  noise synthesizer.
Moreover, while the original DDSP-Add used only $l_{\text{MSSTFT}}$, thus we  used $l_{\text{total}} = l_{\text{MSSTFT}} + l_{f_0}$ for all three in our implementation.
Like SawSing, we set $K=150$ for DDSP-Add.
DWTS only needs small $K'$ as the wavetables are learnable; we set $K'=20$, with wavetable length being $B=512$.

We also employed the neural source-filter (\textbf{NSF}) waveform model \cite{nsf}, which was proposed before the notion ``DDSP'' was coined \cite{ddsp}.
Unlike SawSing, 
NSF uses unweighted sinusoids (i.e., $\sum_{k=1}^{K} \sin(\phi_k(t))$) as the source signal, 
and uses stacked dilated-convolution blocks instead of a simple LTV-FIR filter. 
We adapted the open-source code from the original authors to implement NSF, as well as the following three baselines.

For GAN-based neural vocoders, we used  \textbf{PWG} \cite{pwg} and \textbf{HiFi-GAN} \cite{hifigan}, both of which were  developed for speech, not singing.\footnote{While we did not consider SingGAN \cite{singgan} in the evaluation for lack of open source code, we should have included PeriodNet \cite{periodnet21icassp} for it seems easy to implement. Unfortunately we are aware of PeriodNet too late.} 
PWG is a non-autoregressive version of WaveNet \cite{wavenet} that learns to transform a random noise into target audio with 30 layers of dilated residual convolution blocks, conditioning on the mel-spectrogram. 
For HiFi-GAN, we used the most powerful ``V1'' configuration \cite{hifigan}, which converts a mel-spectrogram into a waveform directly via 12 residual blocks. 
It uses a sophisticated multi-receptive field fusion module in the generator, and multiple multi-scale and multi-period discriminators \cite{hifigan}.

An increasing number of diffusion-based vocoders have been proposed in the past two years for speech \cite{chen2021wavegrad,diffwave,fastdiff,lam2022bddm}.  We adopt as a baseline the   
\textbf{FastDiff} model \cite{fastdiff}, which has been shown to beat HiFi-GAN~V1~\cite{hifigan} and diffusion-based models WaveGrad \cite{chen2021wavegrad} and DiffWave \cite{diffwave} in the mean-opinion-score (MOS) of vocoded speech in listening tests. However, while a noise schedule predictor has been devised to reduce the sampling steps of the denoising Markov chain, the inference time of FastDiff is still around  10 times slower than HiFi-GAN, according to \cite{fastdiff}.

None of these baselines have been trained on MPop600, which is a relatively new dataset. 
Therefore, we trained all these models from scratch with the MPop600 data.

\subsection{Dataset \& Scenarios}  
Our data is from MPop600 \cite{mpop600}, a 
set of accompaniment-free Mandarin singing recordings 
with manual annotation of word-level audio-lyrics alignment. 
Each recording covers the first verse and first chorus of a song.
We used the data from a female singer (named \texttt{f1}) and a male singer (\texttt{m1}); each has 150 recordings. 
For each singer, we reserved 3 recordings (totalling 3.4--3.6 minutes in length) as the \emph{test} set for subjective evaluation, 24 or 21 recordings (around 27--28 minutes) as the \emph{validation} set for objective evaluation, and used the rest (around 3 hours) as the \emph{training} set. 
We trained vocoders for \texttt{m1} and \texttt{f1} independently.

To study the training  efficiency of different approaches, we considered the following two scenarios. We used the same validation and test sets for both scenarios.
\begin{enumerate}[(a)]
    \item \emph{Regular}~[3h~data,~well-trained]: we used the full training data to train the vocoders for each singer for up to 2.5 days (i.e., when the training loss of most vocoders converged), and picked the epoch that reaches the lowest validation loss for each vocoder independently.
    We note that the amount of training time in this ``regular'' scenario is smaller than those seen in speech vocoders \cite{ljspeech}, posing challenges for all the considered models.
    \item \emph{Resource-limited}~[3min data,~3h training]: 
    in a rather extreme case, we randomly picked 3 recordings from the training set (that collectively cover every phoneme at least once) 
    per singer (3.2--3.4 minutes) for training,
    using always the epoch at 3-hour training time. 
\end{enumerate}
For fair comparison, we train the vocoders of different approaches using a dedicated NVIDIA GeForce RTX 3090 GPU each, fixing the batch size to 16 excerpts.

\begin{figure}[t]
\centering
\includegraphics[width=0.49\linewidth]{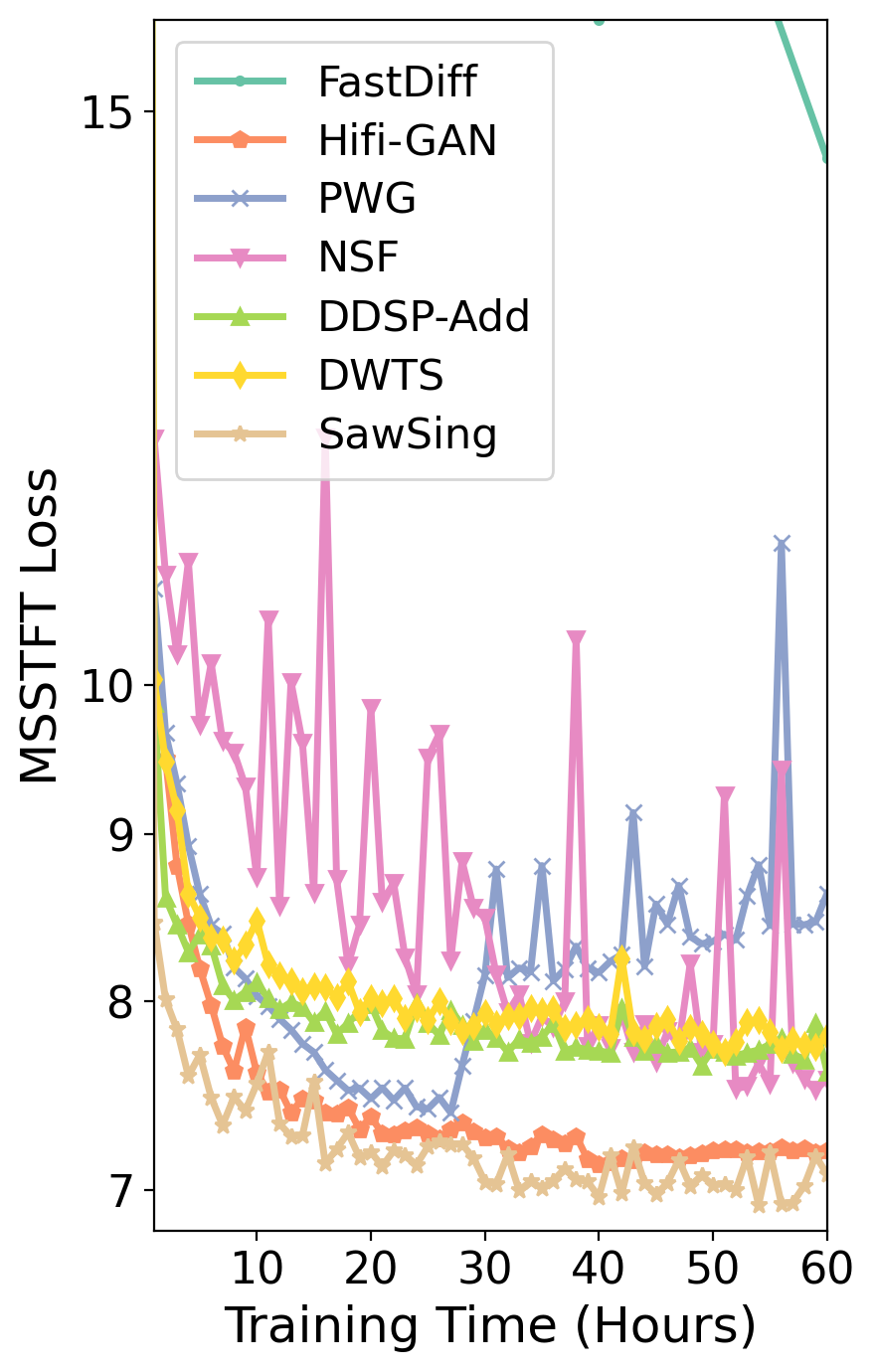}
\includegraphics[width=0.49\linewidth]{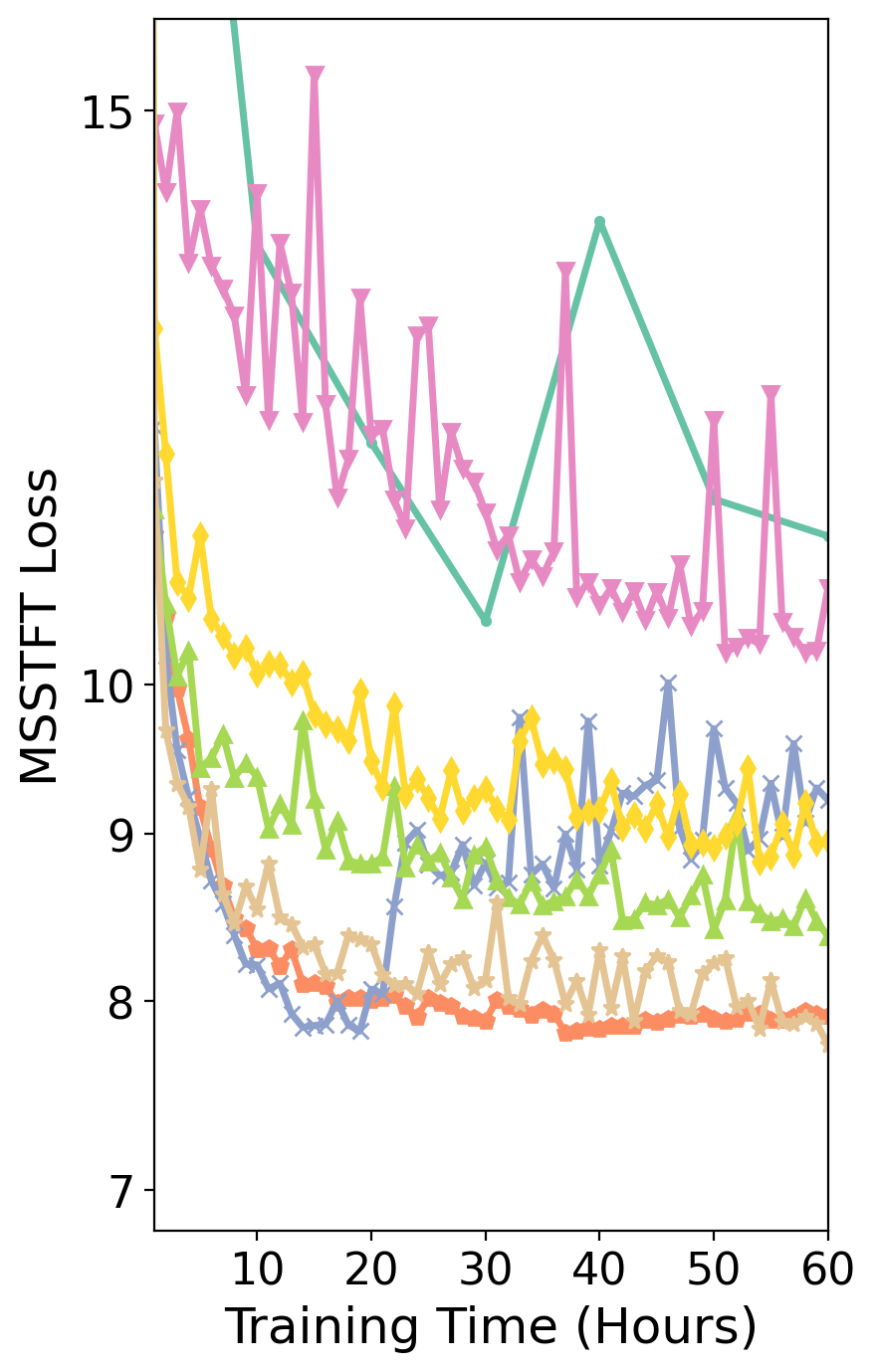}\\
~~~~~~~ (i) female  ~~~~~~~~~~~~~~~~~~~~~~~~~~~~ (ii) male
\caption{The MSSTFT loss on the validation set of different vocoders in the 3-hour data \& well-trained scenario.}
\label{fig:val_loss}
\end{figure}

\section{Objective Evaluation}
\label{sec:result}

For objective evaluation, we reported the MSSTFT and the mean absolute error (MAE) in f0, as well as the Fr\'echet audio distance (FAD) \cite{kilgour2019frechet} 
between the validation data and the reconstructed ones by the vocoders.
FAD measures the similarity of the real data distribution and generated data distribution in an embedding space computed by a pretrained VGGish-based audio classifier, and 
may as such better reflect the perceptual quality of the generated audio.

Figure \ref{fig:val_loss} shows the validation MSSTFT loss as a function of the training time in the regular scenario for the two singers. In Figure \ref{fig:val_loss}(i), SawSing converges faster than the other models and reaches the lowest loss (i.e., 6.93), followed by HiFi-GAN and PWG. While DDSP-add and DWTS converge similarly fast as SawSing, they reach at a slightly higher loss (around 7.50). In Figure \ref{fig:val_loss}(ii), Sawing, HiFi-GAN and PWG perform comparably in the first 20 hrs.
For both singers, PWG overfits when the training time gets too long. 
Moreover, FastDiff converges the most slowly, followed by NSF. Even with 60-hour training time, the MSSTFT of FastDiff remains to be high (e.g., 14.5 for the female singer), suggesting that our training data might not be big enough for this diffusion-based model.\footnote{In the original paper \cite{fastdiff}, FastDiff was trained on 24 hours of speech data from a female speaker \cite{ljspeech}, using 4 NVIDIA V100 GPUs. We tried DiffWave \cite{diffwave} but it converged similarly slow on our data.} 

Table \ref{tab:obj} shows the scores in all the three metrics on the validation set for both scenarios, using the epoch  (a) at the lowest validation loss or (b) at 3h training.
Despite having few trainable parameters, SawSing performs the best in MSSTFT and FAD across both scenarios and both singers, demonstrating its effectiveness as a singing vocoder. 
For scenario (b), DDSP-Add obtains the second-lowest MSSTFT and FAD across the two singers.

For MAE-f0, SawSing attains scores comparable to the best baseline models. 
\yhyang{The average MAE-f0 of SawSing is less than a semitone (100 cents).}
Future work can use a specialized module (e.g., \cite{kim2018crepe}) for the f0 prediction part in $f_\text{NN}$ of SawSing to further improve the MAE-f0. 

Table \ref{tab:obj} also shows that the performance gap between scenarios (a) and (b) in all the three metrics tend to be greater for the diffusion- and GAN-based vocodoers than for NSF and the DDSP-based vocoders.
Besides, among the evaluated models, the performance gap between (a) and (b) is the smallest in the result of SawSing. 
In the resource-limited scenario (b), the FAD of HiFi-GAN reaches only 3.50 and 10.5 for the female and male singers, respectively, while the FAD of SawSing can be  lower than 1.0.
This demonstrates that a strong inductive bias like those employed in NSF and the DDSP-based vocoders is helpful in scenarios with limited training data and training time.

Table~\ref{tab:obj} also displays the real-time factor (RTF) of the models when being tested on a single  NVIDIA 3090 GPU. 
We see that SawSing and DDSP-Add 
have the lowest RTF (i.e., run the fastest),
followed by HiFi-GAN.



According to Table~\ref{tab:obj},  HiFi-GAN performs the best on average among the first three 
vocoders across scenarios and singers. NSF and the DDSP-based vocoders obtain comparable  scores, but DWTS is notably slower.
Hence, we pick HiFi-GAN, NSF, DDSP-Add and SawSing to be further evaluated in the user study below.

\vspace{-2mm}
\section{Subjective Evaluation}
\label{sec:result2}

We conducted an online  study to evaluate the performance of the 4 selected models.
We had 2 sets of questionnaires, one for the female and the other for the male singer.
For each singer, we prepared 8 clips from the 3 \emph{testing} recordings (i.e., totally unseen at training/validation time), each clip corresponding to the singing  of a \emph{full sentence}. 
We let the vocoders trained in scenario (a) to reconstruct the waveforms from the mel-spectrograms of 4 of the clips, and those of (b) for the other 4 clips.
A human subject was requested to use a headset to listen to 5 versions of each of the clip, namely the original `ground truth` recording and the reconstructed ones by the 4 selected models, with the ordering of the 5 versions randomized, the ordering of the 8 clips randomized,  and not knowing the scenario being considered per clip.
The loudness of the audio files were all normalized to --12dB LUFS beforehand using \texttt{pyloudnorm} \cite{pyloudnorm}.
After listening, the subject gave an opinion score from 1 (poor) to 5 (good) in a 5-point Likert scale to rate the audio quality for each audio file. 

\begin{figure}[t]
\includegraphics[width=\linewidth]{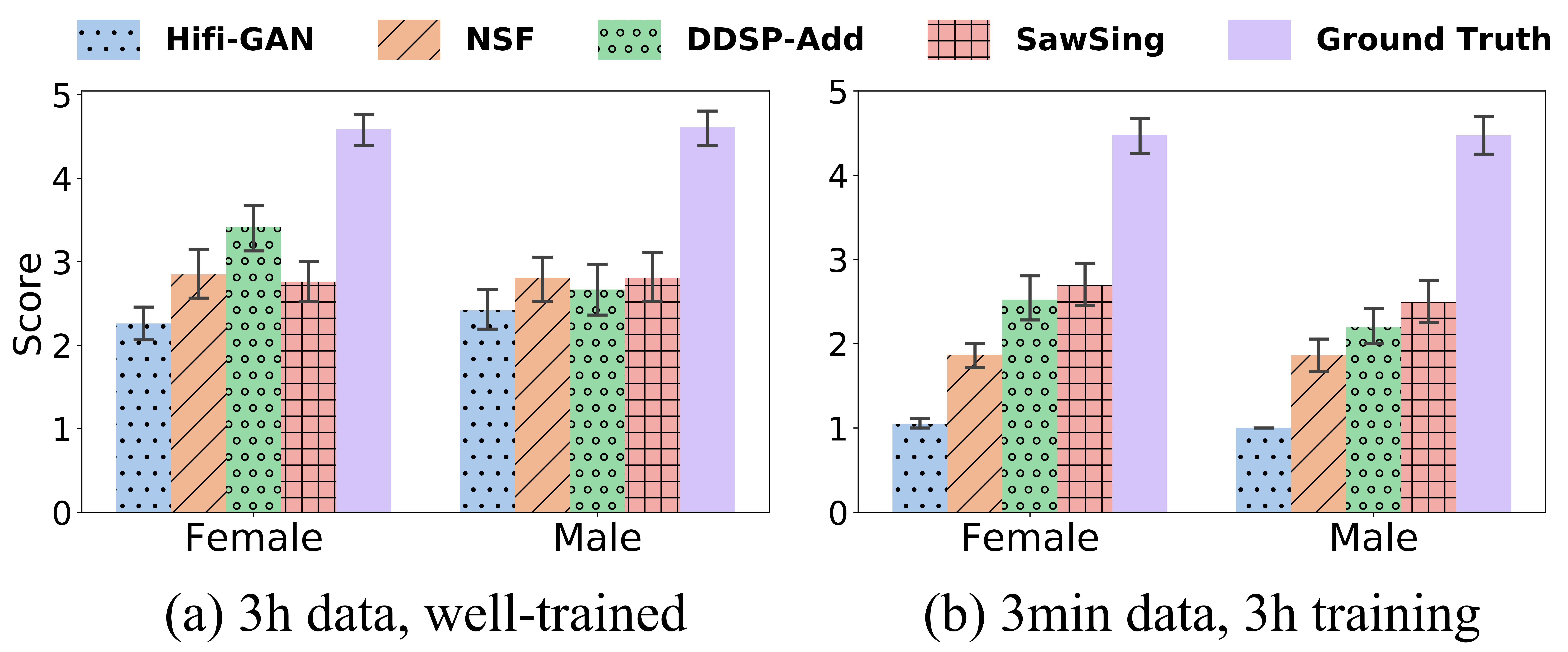}

\caption{MOS with 95\% confidence intervals for subjective evaluation of vocoders trained in the two scenarios.} 
\label{mos}
\end{figure}

Figure~\ref{mos} shows the MOS from 23 anonymized participants for the female and 18 participants for the male singer.
In scenario\,(a), we see that the MOS of the vocoders, including the SOTA HiFi-GAN, mostly reaches 2--3 only, suggesting that training a vocoder on 3-hour data is already challenging. As HiFi-GAN involves a complicated GAN training and much more parameters, its MOS turns out to be significantly lower than those of NSF and the DDSP-based vocoders ($p$-value$<$0.05 in paired t-test). 
Interestingly, while there is no statistical difference among the MOS of NSF, DDSP-Add and SawSing for the male singer in scenario\,(a), DDSP-Add unexpectedly outperforms both NSF and SawSing by a large margin, with statistically significant difference ($p$-value$<$0.05). 

\begin{figure}[t]
\centering
\includegraphics[width=.93\linewidth]{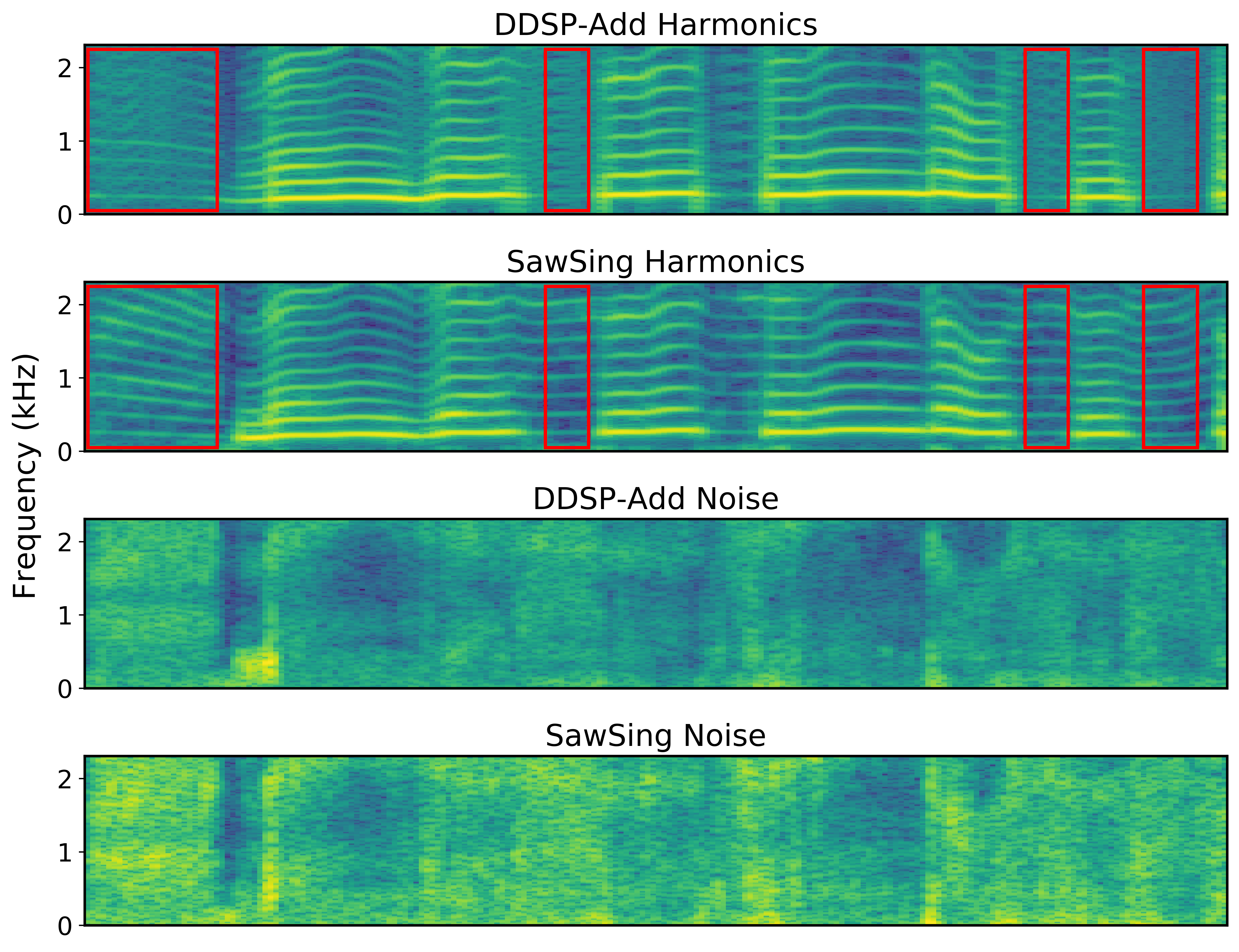}\\
\caption{The spectrograms of the harmonic signal $x_\texttt{h}$ and noise signal $x_\texttt{n}$ generated by DDSP-Add and SawSing for the same  clip. 
The red rectangles highlight the moments the buzzing artifact of SawSing emerges. 
} 
\label{fig:buzz}
\end{figure}

Listening to the result of SawSing 
reveals that its output contains an audible electronic noise, or ``buzzing'' artifact, notably 
when singers emphasize the airflow with breathy sounds 
and for unvoiced consonants such as \texttt{/s/} and \texttt{/t/}.
DDSP-Add is free of such an artifact.
As shown in Figure~\ref{fig:buzz}, 
such artifact appears to due to \emph{redundant} harmonics generated by SawSing that ``connect'' the harmonics of two adjacent phonemes at its harmonic signal $x_\texttt{h}$ for breathing and unvoiced consonants. 
\yhyang{This may be due to the limited capacity of the LTI-FIR filter of SawSing in distinguishing between the nuances of voiced (V) and unvoiced (NV) components during sound transients, modeling a transient even as a harmonic signal.}
Unfortunately, it seems that this artifact cannot be reflected in any training loss functions (and objective metrics) we considered, so the network fails to take it into account while updating the parameters. 
Furthermore, human ears are sensitive to such an artifact, contributing to the lower MOS of SawSing compared to DDSP-Add, despite that SawSing might perform better in other phonemes and long utterances.

Figure~\ref{mos} also shows that the DDSP-based vocoders do outperform HiFi-GAN greatly in the resource-limited scenario (b) with only 3 training recordings, nicely validating the training efficiency of the DDSP-based vocoders.
While the MOS of either DDSP-Add or SawSing is above 2; that of HiFi-GAN is only around 1, i.e., its generation is barely audible.
Moreover, SawSing outperforms DDSP-Add in this scenario for both singers, 
with significant MOS difference for the male singer ($p$-value$<$0.05), \yhyang{though not for the female singer.
This shows that, despite the buzzing artifact, the training efficiency of SawSing can give it an edge over other vocoders in resource-limited applications. }




Inspired by \cite{hifi}, we implement a postprocessing method that uses Parselmouth \cite{parselmouth} to get V/NV flags and sets the harmonic synthesizer amplitudes to zero for the NV portions. This removes much of the artifact (see the demo page). We share the code on our GitHub repo.
Future work can incorporate the V/NV flags at the training phase.

\section{Conclusion}
\label{sec:conclude}

In this paper, we have presented SawSing, a new DDSP-based vocoder that synthesizes an audio via the summation of a harmonic component obtained from filtered sawtooth waves and a stochastic component modeled by filtered noise. 
Moreover, we presented objective and subjective evaluations
\yhyang{complementing the lack of experiments in the recent work of Alonso and Erkut \cite{svs_ddsp}, demonstrating for the first time that both SawSing and DDSP-Add   \cite{ddsp,svs_ddsp}} compare favorably with SOTA general-purpose neural vocoders such as HiFi-GAN \cite{hifigan} and FastDiff \cite{fastdiff} for singing voices
in a regular-resource scenario, and has a great performance margin in a resource-limited scenario.

Future work can improve the harmonic filter of SawSing to resolve the artifact, and use lighter-weight non-causal convolutions \cite{caillon22arxiv} for $f_\text{NN}$ for real-time applications.
We can also implement SawSing as a VST audio plugin for usages in creative workflows and music production \cite{deruty22tismir}.




\section{Acknowledgement}
We are grateful to Rongjie Huang and Yi Ren for sharing with us the code of FastDiff, and Shuhei Imai for helping proofread the paper. We also thank the anonymous reviewers for their constructive feedbacks. Our research is funded by grants NSTC 109-2628-E-001-002-MY2 and NSTC 109-2221-E-007-094-MY3 from the National Science and Technology Council of Taiwan.

\bibliography{ISMIRtemplate}

%
%
%
%
%

\end{document}